\documentclass{iaus}
\usepackage{graphicx}

\title[Distribution of Magnetic Shear] 
{Distribution of Magnetic Shear Angle in an Emerging Flux Region}

\author[Gosain, S.]   
{Sanjay Gosain
}

\affiliation{Udaipur Solar Observatory, Physical Research Laboratory, \\ P. Box No. 198,
Udaipur 313001, Rajasthan, India \\ email: {\tt sgosain@prl.res.in} \\}

\pubyear{2010}
\volume{xxx}  
\pagerange{119--126}
\jname{IAU Symposium 273: Physics of the Sun and Star Spots}
\editors{Debi Prasad Choudhary \& Klaus G. Strassmeier, eds.}
\begin{document}

\maketitle

\begin{abstract}
We study the distribution of magnetic shear in an emerging flux
region using the high-resolution Hinode/SOT SP observations.
The distribution of mean magnetic shear angle across the active
region shows large values near region of flux emergence i.e.,
in the middle of existing bipolar region and decreases while
approaching the periphery of the active region.
\end{abstract}

\section{Introduction}
The non-potentiality of the magnetic field in solar active
regions provides the free-energy needed to fuel the energetic
events like solar flares and CMEs. The non-potentiality may
result either due to the shearing motions of the footpoint or
due to the emergence of pre-stressed magnetic flux from the
convection zone into the photosphere. The study of magnetic
flux emergence in active regions is therefore very important
for understanding the flaring activity. The high-quality vector
magnetograms derived from spectro-polarimetric (SP)
observations by space-based Hinode Solar Optical Telescope
(SOT) (Tsuneta {\it et al.} 2008) are very useful for such
studies. Although, the cadence of such SP observations is not
fast enough to capture the dynamic evolution of magnetic field.
One can, however, study the spatial distribution of magnetic
non-potentiality in such active regions.

One of the important parameters used to characterize the
non-potentiality of the magnetic field is the magnetic shear
angle  (Hagyard {\it et al.} 1984). The magnetic shear angle
$\Delta\psi$ is defined as the angle between the azimuths of
the observed and the potential field i.e.,
$\Delta\psi=\psi_{o}-\psi_{p}$. The larger is the magnetic
shear angle (henceforth MSA) more non-potential is the magnetic
field.

The distribution of MSA in an active region, which is going
through flux emergence process, is useful in order to
understand the buildup of non-potentiality in active regions.
In the present paper we present results from such a study.

\section{Observations and Analysis Methods}
The magnetic flux was emerging in the middle of an existing
bipolar region NOAA 10978, during 13 December 2007. Several
thin elongated fibril structures were seen in the region of
flux emergence. These emerging flux tubes displace the
photospheric plasma, causing an appearance of diverging flow
pattern. As the flux emergence progresses the foot points of
the fibrils move apart and become more and more vertical and
bundle together to form sunspots.

The spectropolarimetric observations of this active region were
obtained by the spectropolarimeter instrument (Ichimoto {\it et
al.} 2008) onboard Hinode space mission (Kosugi et al 2007).
The magnetic field vector was derived by using a
 Stokes inversion code called MERLIN (Lites {\it et al.} 2007). The code performs
a non-linear least squares fit of the observed Stokes profiles
with theoretical Stokes profiles computed under the
Milne-Eddington model atmosphere assumptions.

The top panel of figure 1 shows the longitudinal magnetic field
map of this active region  at two different stages during its
evolution. The small bipoles near the polarity inversion line
(PIL), with opposite polarity orientation as compared to the
overall active region polarity suggests the presence of
undulatory U-loops. The bottom panel of figure 1 shows the
continuum intensity maps of the active region. The typical
features of an emerging flux region, like elongated thin fibril
like structures, can be clearly noticed. We evaluate the
variation of shear angle across the active region by computing
the mean shear angle within each of the rectangular box labeled
1-25. The boxes are distributed along the bipolar axis of the
active region so as to get spatial profile of shear across the
active region. We remove the tilt of the active region by
rotating the map so that the bipolar axis is horizontal. The
bipolar axis is computed by joining the centroid of the
longitudinal flux in either polarity of the active region. The
shear angle is computed for each pixel of the map and is
averaged inside each of the box 1-25. The plot in figure 2
shows the distribution of mean magnetic shear angle across the
active region.

\section{Results and Discussions}
It can be seen from figure 2 that the mean shear angle is not
uniformly distributed across the active region. The mean shear
angle is higher in the central portion of the active region
where the flux tubes emerge. The peak value of the  shear angle
in the central portion is about 30-35 degrees, while at the
periphery it is about 25 degrees. The pattern persists during
the two time intervals which are separated by about six hours
as shown in figure 1.

These observations show that the magnetic field in the middle
portion of the emerging flux region consists of the largest
amount of non-potentiality. The present results obtained with
the high-resolution and high-sensitive polarimetry put shear
distribution studies on a firm footing. Similar results were
obtained by Schmieder {\it et al.} (1996) for NOAA 6718 where
the comparison of coronal field structure with linear
force-free field computation yielded a differential magnetic
field shear model. The present observations support their
argument that the decreased shear in the outer portions of the
active region is probably due to continual relaxation of the
magnetic field to lower energy state in the older portions of
the active region. In future we plan to carry out a similar
analysis of magnetic shear in decaying or dormant active
regions to see how different the pattern appears as compared to
the emerging flux regions.

\section{Acknowledgements}
The presentation of this paper in the IAU Symposium 273 was
possible due to financial support from the National Science
Foundation grant numbers ATM 0548260, AST 0968672 and NASA -
Living With a Star grant number 09-LWSTRT09-0039. Hinode is a
Japanese mission developed and launched by ISAS/JAXA, with NAOJ
as domestic partner and NASA and STFC (UK) as international
partners. It is operated by these agencies in co-operation with
ESA and NSC (Norway).

\begin{figure}    
\centerline{\includegraphics[width=1.0\textwidth,clip=]{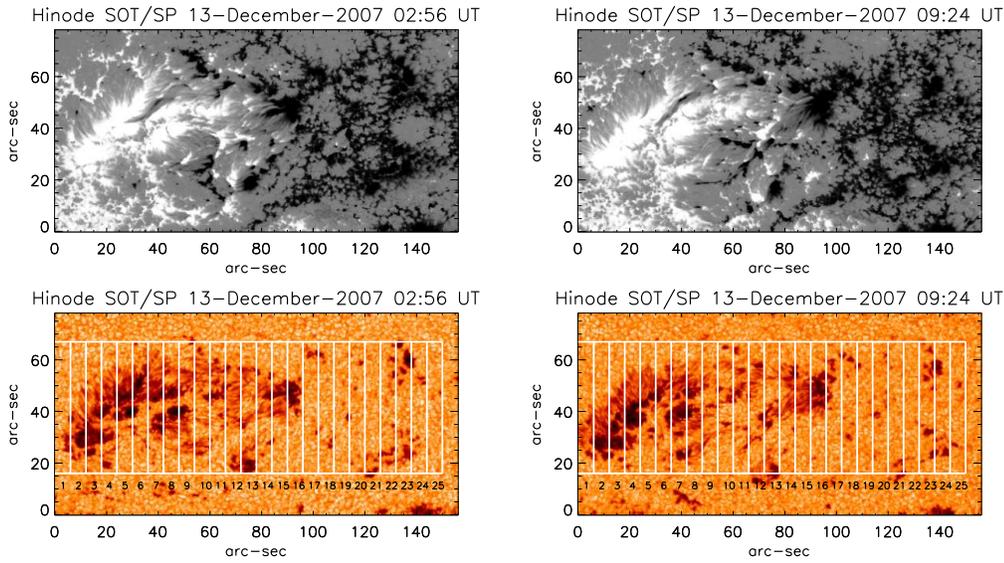}}
\caption{The top panels show the longitudinal field map of the emerging flux region in NOAA 10978 during 13 December 2007. The bottom
panels show the continuum intensity maps of this region. The distribution of magnetic shear across the region is estimated by calculating the
mean shear angle inside rectangular boxes labeled 1 through 25. The variation of magnetic shear angle across the region from boxes 1 to 25 is shown in figure 2.}
\end{figure}

\begin{figure}    
\centerline{\includegraphics[width=.45\textwidth,clip=]{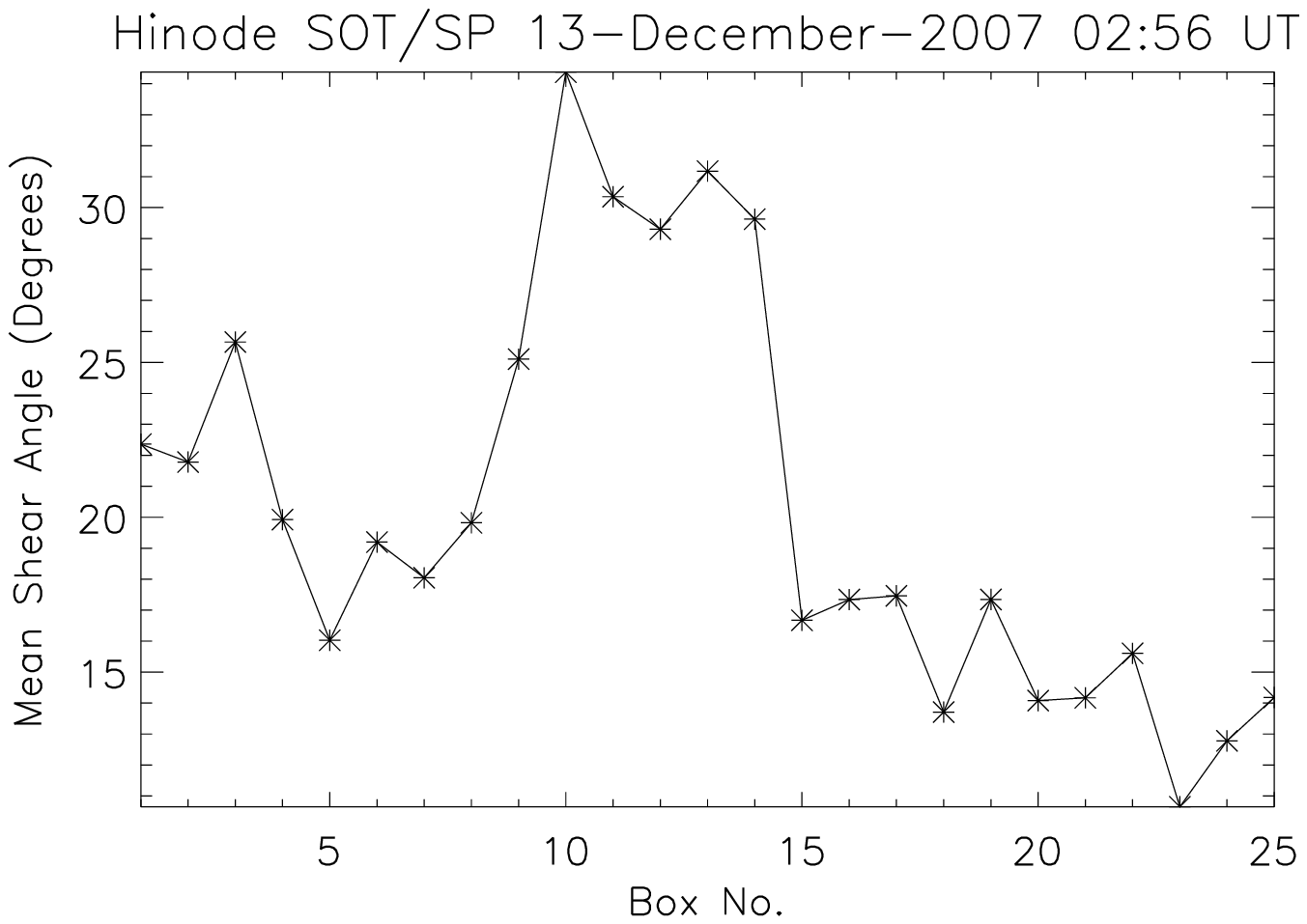}\hspace{0.25in}\includegraphics[width=.45\textwidth,clip=]{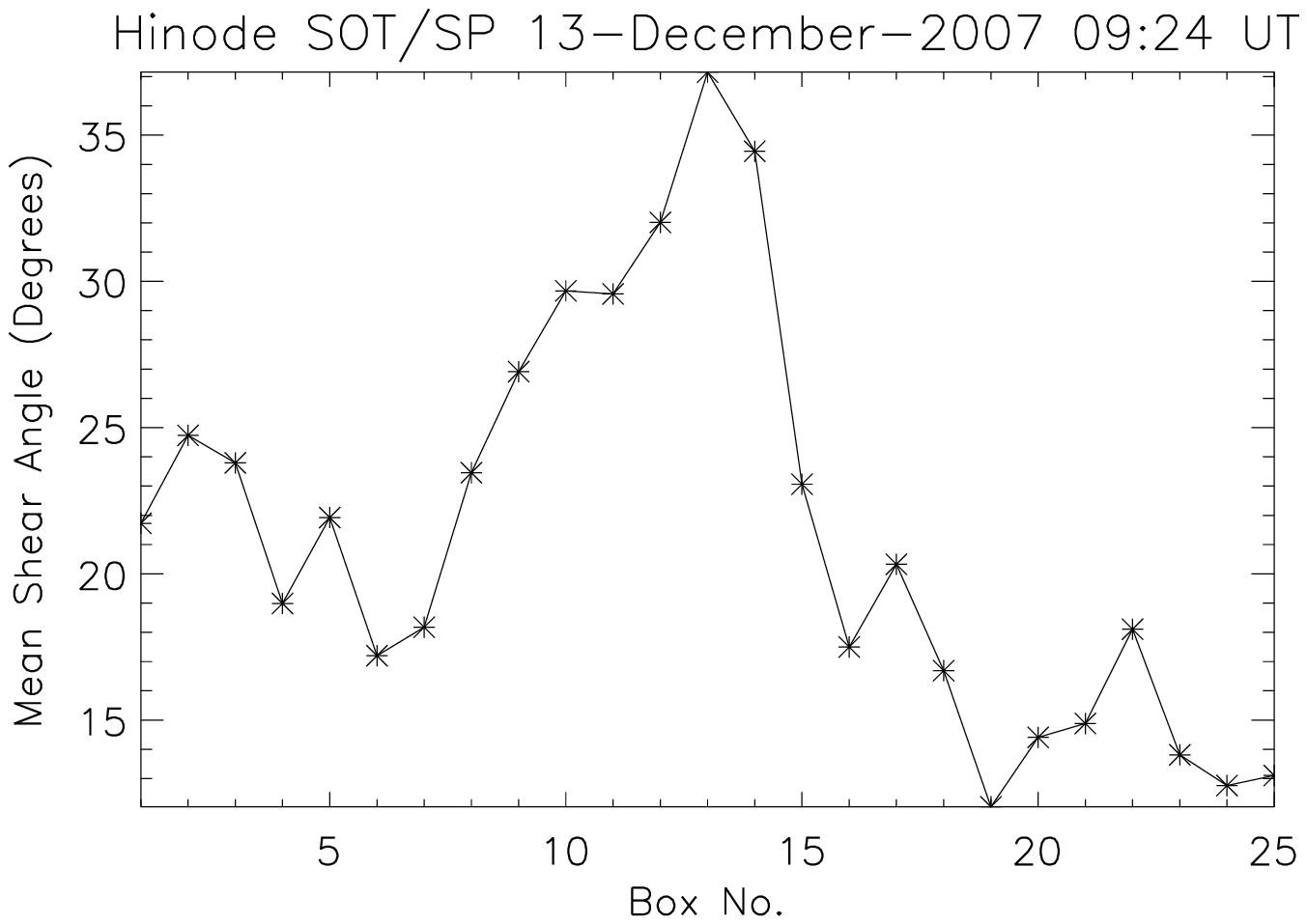}}
\caption{The left and right panels show the distribution of magnetic shear angle across the active region at two different times during 13 December 2007.
The abscissa shows the box number and the ordinate shows the mean shear angle inside the corresponding box.    }
\end{figure}

\end{document}